\def\version{KIAS-P98015}
\documentstyle[11pt,epsfig]{article}
\def\bea {\begin{eqnarray}}
\def\eea {\end{eqnarray}}
\def\be {\begin{equation}}
\def\ee {\end{equation}}
\def\ben {\begin{enumerate}}
\def\een {\end{enumerate}}

\def\calO{{\cal O}}
\def\calM{{\cal M}}

\def\bfp{{\bf p}}

\def\bfq{{\bf q}}

\def\bfr{{\bf r}}

\newcommand{\e}{{\mbox{e}}}

\def\del{\partial}

\def\roughly#1{\mathrel{\raise.3ex\hbox{$#1$\kern-.75em%
\lower1ex\hbox{$\sim$}}}}
\def\lsim{\roughly<}
\def\gsim{\roughly>}
\def\fm{{\mbox{fm}}}
\def\MeV{{\mbox{MeV}}}
\def\erf{{\mbox{erf}}}
\def\erfc{{\mbox{erfc}}}

\renewcommand{\thefootnote}{\#\arabic{footnote}}
\begin{document}
\begin{center}
\begin{flushright}
  \version
\end{flushright}
\vskip 1cm
{\Large \bf The Power Of Effective Field Theories In Nuclei:\\
 The Deuteron, $NN$ Scattering and Electroweak Processes}
\vskip 0.5cm
{{\large Tae-Sun Park}\footnote{
  Corresponding auther.
  e-mail: park@alph02.triumf.ca,
  tel.: 1-604-222-1047, 
  fax: 1-604-222-1074,\\
  postal address: TRIUMF, Wesbrook Mall,
  Vancouver, B. C., V6T 2A3, Canada} \\

{\it Grupo de F\'\i sica Nuclear,
  Universidad de Salamanca, 37008 Salamanca, Spain} \\
{\it and}\\
{\it School of Physics, Korea Institute for Advanced Study,
 Seoul 130-012, Korea} \\
\vskip 2mm
 {\large Kuniharu Kubodera}\footnote{
  e-mail: kubodera@nuc003.psc.sc.edu}
\\
{\it Department of Physics and Astronomy, University of South Carolina\\
   Columbia, SC 29208, U.S.A. }\\
\vskip 2mm
{\large  Dong-Pil Min}\footnote{
  e-mail: dpmin@phya.snu.ac.kr}
\\
{\it Department of Physics and Center for Theoretical Physics \\
  Seoul National University, Seoul 151-742, Korea} \\
\vskip 2mm
 {\large Mannque Rho}\footnote{
  e-mail: rho@spht.saclay.cea.fr}
\\
{\it Service de Physique Th\'eorique, CEA Saclay \\
  91191 Gif-sur-Yvette Cedex, France}
}
\end{center}

\vfill{
\centerline{\bf Abstract}
\vskip 0.1cm

We show how {\it effectively} effective quantum field theories work
in nuclear physics. Using the physically transparent cut-off
regularization, we study the simplest nuclear systems of two
nucleons for both bound and scattering states at a momentum scale much less
than the pion mass. We consider all the static properties of the deuteron,
the two-nucleon scattering phase-shifts, the $n+p\rightarrow d+\gamma$
process at thermal energy and the solar proton fusion process
$p+p\rightarrow d+e^+ +\nu_e$, and we demonstrate
that these are all described with great accuracy
in the expansion to the next-to-leading order.
We explore how a ``new" degree of freedom enters in an effective theory
by turning on and off the role of the pion in the Lagrangian.
}

\newpage

\renewcommand{\thefootnote}{\#\arabic{footnote}}
\setcounter{footnote}{0}
\section{Introduction}
\indent

The idea of effective quantum field theory\cite{effective,pol,lepageTASI}
is extremely simple. At low energy where non-perturbative effects of
QCD dominate, the relevant degrees of freedom are not quarks and
gluons but hadrons. Let the relevant hadronic degrees of freedom be
represented by the generic field $\Phi$. Separate the field into
two parts; the low-energy part $\Phi_L$ in which we are interested
and the high-energy part $\Phi_H$ which does not interest us
directly. We delineate the two parts at the momentum scale characterized
by a cutoff, say, $\Lambda_1$.
In the generating functional $Z$
that we want to compute, integrate out the $\Phi_H$ field and
define an effective action $S^{eff}_{\Lambda_1}[\Phi_L] =\frac{1}{i}\ln
\left(\int [d\Phi_H]e^{iS[\Phi_L,\Phi_H]}\right)$.
Then what we need to compute is
\bea
Z=\int_{\Lambda_1} [d\Phi_L] e^{iS^{eff}_{\Lambda_1} [\Phi_L]}.
\eea
Given the right degrees of freedom effective below the cutoff
$\Lambda_1$, we expand the effective action as
$S^{eff}_{\Lambda_1}[\Phi_L]=\sum_i^\infty C_i Q_i[\Phi_L]$,
where $Q_i[\Phi_L]$ are local field operators allowed by the symmetries
of the problem,
and $C_i$'s are constants,
which are required to satisfy  the ``naturalness condition".
The $Q_i[\Phi_L]$'s are ordered in such a manner that,
for low-energy processes,
the importance of the $Q_i[\Phi_L]$ term diminishes as $i$ increases.
The contribution of the $\Phi_H$ degrees of freedom
integrated out from the action is not simply discarded but instead gets
lodged in the coefficients $C_i$  as well as in higher-dimensional operators.
The strategy of effective field theories is to truncate the series
at a manageable finite order, and obtain desired accuracy to capture the
essence of the physics involved.
The separation of $\Phi$ into $\Phi_L$ and $\Phi_H$
involves two steps: The first is to ``decimate'' the degrees of freedom,
that is, to choose the fields $\Phi_L$
relevant for the physics to be incorporated explicitly into the Lagrangian,
and the second is a regularization procedure which
regulates the high-momentum contribution.
Let $m$ be the lightest mass of the degrees of freedom
that are integrated out.
Clearly the most physically transparent regularization
is to introduce a momentum cutoff with $\Lambda_1 \sim m$.
A different cutoff introduces different coefficients $C_i$ for the
given set of fields.
The idea is then to pick a value of the cutoff
that is low enough to avoid unnecessary complications
but high enough to avoid throwing away relevant physics.

In the full (un-truncated) theory, the location of the cutoff is entirely
arbitrary, so we could have chosen
any values of the cutoff other than $\Lambda_1$
without changing physics.
The cut-off dependence of $Q_i[\Phi_L]$ is compensated by
the corresponding changes in $C_i$ through the
renormalization group equation.
In such a theory,
which is rarely available in four dimensions,
physical observables should
be strictly independent of where the cutoff is set.
This of course will not be the case when the series is truncated.
However,
for a suitably truncated effective theory to be predictive,
the observables should be more or less insensitive
to where the cutoff is put.
Should there be strong cutoff dependence, it would signal
that there is something amiss in the scheme.
For instance,
it could be a signal for the presence of new degrees of
freedom or of ``new physics" that must be incorporated and/or
for the necessity of including higher-order terms.
How  ``new physics" enters in our case will be studied in terms of the role
that the pion -- a ``new degree of freedom" -- plays at very low energy.
Here we of course know exactly what the ``new physics" is but
there will be a lesson to learn from this exercise in exploring beyond
``standard models". The following ``toy model" illustrates our point.
Let us imagine that we knew that QCD is the correct microscopic theory for
the strong interactions with chiral symmetry realized in the Nambu-Goldstone
mode but the pions were not  yet discovered and that the only ``effective"
degrees of freedom we had at hand were the proton and the neutron.
Now the question to ask is how to set up a systematic theory that could
describe very low-energy properties of the deuteron, the NN scattering
and the electroweak matrix elements and in doing so how to infer that a
``new degree of freedom" (pion) lurked at $\sim 140$ MeV.
We will see in Section 6 how this question could be studied
in the real low-energy world of nuclei.

An important aspect of effective field theory is that the result
should not depend upon what sort of regularization one uses. In
principle, therefore, one could equally well use the dimensional
regularization or the cut-off regularization or something else as
long as one makes sure that all relevant symmetries are properly
implemented~\footnote{There are of course certain issues which
raise technical, though not fundamental, problems, e.g., how to
preserve chiral symmetry in the cut-off procedure and how to
correctly handle short-distance physics in the standard dimensional
regularization etc. For a recent discussion on this matter, see
\cite{donoghue}.}.
Calculated to high enough order in a consistent expansion, the results
should in principle be the same up to that order
although in practice there is a
certain amount of art involved in the choice of the scheme.
This issue in the context of the dimensional regularization and its
variants was addressed recently in \cite{gegelia2}. Here we will
use the cut-off regularization for the reason that the procedure is
simple and physically transparent~\cite{lepage,donoghue} when
limited to the order we shall adopt, namely, to the next-to-leading
order (NLO). For some of the problems treated here such as
neutron-proton ($np$) scattering,
one could also use in a predictive manner a modified 
dimensional regularization in which certain power divergences are 
removed (i.e., PDS scheme)\cite{KSWPDS}.
For further comments on the PDS scheme, see 
Gegelia~\cite{gegelia2}.
It is not yet fully clear to us 
how such a scheme despite its advantages could avoid some of
the problems raised in \cite{gegelia2,donoghue,lepage}.

In this paper, we show that an effective field theory with finite cut-off
regularization works surprisingly well for
all low-energy observables in the two-nucleon 
bound and scattering states;
e.g., the static properties of the deuteron,
the low-energy $np$ scattering, the
radiative $np$ capture, $n+p\rightarrow d+\gamma$,
at thermal energy and the proton fusion, $p+p\rightarrow d+e^++\nu_e$,
in the Sun.
Some of the results were reported in previous
publications~\cite{pmr,pkmr-prc,pkmr-apj} but they will be included
here both for the sake of completeness and
for presenting our more recent sharpened thoughts on the matter.
One of the significant results of this
paper is a quantitative justification from ``first principles" of
the previously formulated ``hybrid method" \cite{pmr} for the
radiative $np$ capture process $n+p\rightarrow d+\gamma$. We justify
quantitatively the procedure of using the
high-quality phenomenological
two-nucleon wavefunctions\footnote{The latest phase-shift
analyses of the two-nucleon data and the construction
of phenomenological potentials to reproduce
the resulting phase shifts
have reached such a level that all the existing data
below $\sim$350 MeV can be reproduced
with high precision\cite{nijmegen,v18}.
We refer to these state-of-the art nucleon-nucleon potentials
as high-quality phenomenological potentials.
The two-nucleon wave functions generated
by these potentials are referred to as
high-quality phenomenological wave functions.}
supplied by the Argonne $v_{18}$
potential~\cite{v18} in calculating the matrix element of the
electroweak operators derived by chiral perturbation
theory with ``irreducible" diagrams, a procedure first suggested
by Weinberg~\cite{weinberg} (for an early implementation, see
\cite{vankolck}).
Similar support
is given for the
hybrid-method calculation
based on the high-quality phenomenological wavefunctions
of the proton fusion process $p+p\rightarrow
d+e^++\nu_e$ crucial in the solar neutrino problem.
This constitutes an important application of effective theories
for nuclei to astrophysics.

\section{Formulation of the Problem}
\indent

The problem we shall address is the following. Suppose we wish to
calculate, starting from the ``fundamental principles" of QCD,
the structure of the deuteron and the scattering of two nucleons at
very low energy, with the characteristic momentum probed $p$ much less
than the pion mass $m_\pi\approx 140$ MeV. What are the relevant
variables? Quarks and gluons are clearly not the {\it right}
variables. We introduce instead the nucleon field $\psi$ in the
Lagrangian as a matter field~\footnote{In principle we could start
with a Lagrangian consisting of the Goldstone meson fields only,
obtain the baryons as solitons (``skyrmions") and then proceed to
compute the fluctuations. But this line of research has not yet met
with  a quantitative success.  See \cite{NRZ} for a discussion on
this matter.}. Next we have the option of having or not having
the pion field figure explicitly in the Lagrangian. In the real world,
the pion has a mass of about 140 MeV, much larger than the momentum scale
we are interested in but in the chiral limit, it would be
degenerate with the vacuum.
So one might ask whether
it is necessary to retain the pion in the framework
even when the pion mass is taken into account.
To address this question, we will work to the
next-to-leading order (NLO) in the momentum expansion a la
Weinberg~\cite{weinberg}.\footnote{Recently, Weinberg's counting
rule (\ref{nuChPT})
has been questioned by the authors of \cite{KSWPDS}
in connection with the large scattering length for the
S-wave $np$ scattering. Our use of the Weinberg counting is not affected
by this criticism. This matter is discussed in Appendix.}
It is then sufficient to take the
two-nucleon potential written to NLO and insert it into an infinite
series of reducible graphs to be summed -- which is equivalent to
solving the Schr\"odinger or Lippman-Schwinger equation
with that potential.

For the sake of completeness of presentation, we first describe the
general features of the counting rules adopted in this article. The
way these counting rules are implemented in our formulation will
become clear as we describe how the terms that enter into the
observables are computed.

For processes whose typical momentum scale can be characterized by
the pion mass, the pion field should figure explicitly in the
Lagrangian. With nucleons treated as ``heavy" and the anti-nucleon
degrees of freedom integrated out, one obtains a consistent and
systematic expansion scheme
 -- heavy-baryon chiral perturbation theory --
where nucleons and pions are the dynamical fields.
Let $Q$ be the typical momentum scale of the process or the pion mass,
which is regarded as ``small" compared to the chiral scale
$\Lambda_\chi \sim 1$ GeV.
The counting rule given by Weinberg\cite{weinberg} is that,
a Feynman graph comprised of $A$ nucleons,
$N_E$ external (electroweak) fields,
$L$ {\it irreducible} loops and $C$-separated pieces
is order of $\calO(Q^\nu)$ with
\bea
\mbox{ChPT:}\ \ \
\nu &=& 2 L + 2 (C-1) + 2 - (A+N_E) + \sum_i \bar \nu_i,
\label{nuChPT}\\
\bar \nu_i &\equiv& d_i + \frac{n_i}{2} + e_i -2
\label{barnu}\eea
where we have characterized each vertex $i$
by $d_i$ the number of derivatives and/or the power of the pion
mass, $n_i$ that of nucleon fields and $e_i$ that of external
(electroweak) fields. The quantities $\bar \nu_i$ and $C$ are
defined so that $\bar \nu_i\ge 0$ even in the presence of external
fields \cite{rho91} and $(C-1)=0$ for connected diagrams.

Now if we limit ourselves  to a very low-momentum regime, say,
$p \ll m_\pi$, then even
the pion field can be integrated out and we are left only
with the nucleon field.
Since we have integrated out the anti-nucleon degrees of freedom
as well as all other meson fields, there are no irreducible loops
and the corresponding counting rule can be
obtained simply by putting the number of loops $L$ equal to zero,
\bea
\mbox{EFT:}\ \ \
\nu = 2 (C-1) + 2 - (A+N_E) + \sum_i \bar \nu_i
\eea
with the same definition for $\bar \nu_i$ given in eq.(\ref{barnu}).
But the meaning of $d_i$ in eq.(\ref{barnu}) is changed: Since the
pion field is integrated out, the $d_i$ stands here only for the
number of derivatives, and not the power of the pion mass. Then, up
to NLO, the resulting potential consists of contact interactions
and their two spatial derivatives.

Electromagnetic interactions have different counting rules
and should be treated separately.
Since the Coulomb interaction (between protons) is given as
$\frac{\alpha}{\bfq^2}$ (where $\alpha\simeq 1/137$ is the fine-structure
constant and $\bfq$ is the momentum transferred),
it is of relevance in an extremely small momentum region
while it becomes irrelevant in other regions due to
the smallness of $\alpha$.
Since we are interested in the proton fusion at
threshold, we will explicitly include the Coulomb potential
in the $pp$ sector.

If we increase the typical probe momentum
to a value close to but lower than the pion mass,
the pion degree of freedom will become ``marginal"
and its explicit presence will improve the convergence of the series.
In this article, we will perform an
NLO calculation both with and without the pion field,
thereby revealing its interplay in the effective field theory
(EFT). To this order, pion loops do not enter and hence we can
simply focus on the (irreducible) potential.
Our {\em unregulated} potential ${\cal V}$ has the form
\bea
{\cal V}(\bfq) &=&
- \tau_1\cdot\tau_2 \,\frac{g_A^2}{4 f_\pi^2}\,
\frac{ \sigma_1\cdot \bfq\,\sigma_2\cdot \bfq}{\bfq^2+m_\pi^2}
+ \frac{4\pi}{M} \left[C_0 + (C_2 \delta ^{ij} + D_2 \sigma^{ij})
 q^i q^j \right]
+ Z_1 Z_2 \frac{\alpha}{\bfq^2}
\label{Vq}\eea
with
\begin{equation}
\sigma^{ij} = \frac{3}{\sqrt{8}} \left(
\frac{\sigma_1^i \sigma_2^j + \sigma_1^j \sigma_2^i}{2}
- \frac{\delta^{ij}}{3} \sigma_1 \cdot \sigma_2 \right),
\end{equation}
where $\bfq$ is the momentum transferred, $g_A \simeq 1.26$ the
axial-vector coupling constant, $f_\pi\simeq 93\ \mbox{MeV}$ the
pion decay constant, $M\simeq 940\ \MeV$ the nucleon mass,
and $Z_1 Z_2=1$ for
the $pp$ channel and zero otherwise.
For the proton fusion process,
we will also consider $\calO(\alpha^2)$ corrections, i.e., the
vacuum polarization (VP) potential and the two-photon-exchange
(C2) potential;
as in \cite{pkmr-apj}, these will be treated perturbatively.
The parameters $C$'s and $D_2$ are defined for each channel.
The $D_2$ term, however, is effective only for the spin-triplet
channel. Thus, there are two parameters ($C_0$ and $C_2$) for each of
the $pp$ and $np$ ${}^1S_0$ channels, and three ($C_0$, $C_2$ and
$D_2$) for the ${}^3S_1$ channel.

It is perhaps useful to explain how the potential
(\ref{Vq}) is to be understood. The first (nonlocal) term is the
pion exchange involving the Goldstone boson and hence
completely known.
The (local) terms in the square brackets represent the effect
inherited from the degrees of freedom that have been integrated out,
for instance the heavy mesons whose masses are much
higher than the cutoff. It should be noted that when the pion is
also integrated out, its effect will appear in
the local terms, modifying the constants.
This means that even if we drop the first term of
(\ref{Vq}), we still have part, if not most, of pionic effects in
the theory.

Inserted into an infinite
bubble sum required to describe bound and quasi-bound states,
the potential (\ref{Vq}) will generate an infinite series of
divergent terms unless the integrals are suitably regularized.
This is expected in an effective theory which is not renormalizable in
the conventional (Dyson) sense. As announced we shall regularize the
divergences with a momentum cutoff.  In
\cite{pkmr-prc} where only nucleonic contact terms and their
derivatives were taken into consideration, a  regularization
that preserves the separability of the potential was used. Of course, the
precise form of the regulator should not matter~\cite{lepage}. Since
both the Coulomb interaction and the one-pion-exchange potential
are non-separable, we shall adopt a slightly different
form of the regularization:
in coordinate space, 
\bea
V(\bfr) &\equiv& \int\! \frac{d^3\bfq}{(2\pi)^3}
\, \e^{i\bfq\cdot \bfr}\, S_\Lambda(\bfq^2)\,
 {\cal V}(\bfq),
\label{regV}\eea
where $S_\Lambda(\bfq^2)$ is the regulator with a cutoff $\Lambda$.
For our purpose it is convenient to take the Gaussian
regulator\cite{pkmr-prc,cohen,lepage},
\be
S_\Lambda(\bfq^2) = \exp\left( - \frac{\bfq^2}{2 \Lambda^2} \right).
\label{Sdef}\ee
Inserting eqs.(\ref{Vq}, \ref{Sdef}) into eq.(\ref{regV}), we
obtain
\bea
V(\bfr) &=&
\alpha_\pi\, \frac{\tau_1\cdot\tau_2}{3}\, \left[
 \sigma_1 \cdot \sigma_2\, v_\Lambda(r, m_\pi)
 + S_{12}(\hat r) \, t_\Lambda(r, m_\pi) \right]
\nonumber \\
&+& \frac{4\pi}{M} \left[
C_0' - C_2 \nabla^2_\bfr - \frac{S_{12}(\hat r)}{\sqrt 8} D_2
 r\frac{\del}{\del r} \frac{1}{r} \frac{\del}{\del r}
 \right] \delta^3_\Lambda(\bfr)
\nonumber \\
&+&
 Z_1 Z_2 \alpha \, v_\Lambda(r, 0)
\label{Vr}\eea
with $S_{12}(\hat r) \equiv 3 \sigma_1\cdot{\hat r} \,\sigma_2\cdot
 {\hat r} - \sigma_1\cdot \sigma_2$,
\bea
\delta_\Lambda^3(\bfr) &=&
\int\! \frac{d^3\bfq}{(2\pi)^3}
\, \e^{i\bfq\cdot \bfr}\, S_\Lambda(\bfq^2)\,,
\\
v_\Lambda(r, m) &=& \frac{4\pi}{S_\Lambda(-m^2)} \,
\int\! \frac{d^3\bfq}{(2\pi)^3} \, \e^{i\bfq\cdot \bfr}\,
\frac{S_\Lambda(\bfq)}{\bfq^2 + m^2},
\\
t_\Lambda(r, m) &=&
\frac{r}{m^2} \frac{\del}{\del r} \frac{1}{r} \frac{\del}{\del r}
v_\Lambda(r, m)
\eea
and
\be
\alpha_\pi \equiv
 S_\Lambda(-m_\pi^2) \frac{g_A^2 m_\pi^2}{16 \pi f_\pi^2}.
\ee
The factor $S_\Lambda^{-1}(-m_\pi^2)$ in the definition of
$v_\Lambda(r, m)$ is introduced so that $v_\Lambda(r, m)$ and
$t_\Lambda(r, m)$ become Yukawa functions in the large-$r$ limit:
\bea
\lim_{\Lambda r \gg 1} v_\Lambda(r, m) &=& \frac{\e^{- m r}}{r},
\nonumber \\
\lim_{\Lambda r \gg 1} t_\Lambda(r, m) &=&
 \left(1 + \frac{3}{m r} + \frac{3}{(m r)^2} \right)
 \frac{\e^{- m r}}{r}.
\eea
In eq.(\ref{Vr}), we have absorbed the $\delta_\Lambda^3(\bfr)$
term appearing in the one-pion-exchange potential into the nucleon contact
interactions by redefining $C_0$ as
\be
C_0' \equiv C_0 - \frac{M \alpha_\pi}{3 m_\pi^2 S_\Lambda(-m_\pi^2)}
\tau_1\cdot\tau_2\,
 \sigma_1 \cdot \sigma_2.
\ee
Hereafter we will omit the prime on $C_0$.
As for the one-pion-exchange potential,
we set $\alpha_\pi= 0.075$ following \cite{v18,nijmegen}.\footnote{
Note that our EFT potential (\ref{Vq}) resembles
the potential used in Nijmegen group's multi-energy analysis \cite{nijmegen}.
They used -- in fitting their potential -- 1787 $pp$ and 2514 $np$
accurate scattering data below $T_{\rm lab}=350$ MeV
published between 1955 and 1992.
They separated their potential into the long-range potential $V_L$ and
the short-range potential $V_S$ such that $V(r) = V_L(r)$ for $r \ge b$,
and $V(r) = V_S(r)$ otherwise, with $b= 1.4$ fm.
Their long-range potential consists of electromagnetic interactions
and one-pion-exchange potentials, as well as heavy-meson-exchange
(such as $\omega$, $\rho$ and $\eta$) potentials,
but since $V_L$ is applied only for $r \ge b$,
the heavy-meson exchange is irrelevant except for achieving high precision
in the large momentum region.
The short-ranged part is parameterized by a square-well
potential (of radius $b$) with an $E$-dependent depth.
The $E$-dependence is given  by a finite series
in increasing powers of $E$. Taking $1\sim3$ terms for
each channel (except for the $pp$ ${}^1S_0$
channel for which four terms are retained),
they achieve a very accurate fit with $\chi^2/\mbox{data} = 1.08$.}

Let us write the potential for each channel of interest.
\vskip 0.3cm

$\bullet$
{\bf The $np$ and $pp$ $^1S_0$ channel}:
Here tensor interactions are ineffective and so
the potential reduces, with eq.(\ref{dd}), to
\bea
V(\bfr; {}^1S_0) &=&
 - \alpha_\pi\, v_\Lambda(r, m_\pi)
+ \frac{4\pi}{M} \left[ C_0^{\rm 1S0} -
 \Lambda^2 (\Lambda^2 r^2 -3 )\, C_2^{\rm 1S0}
 \right] \delta^3_\Lambda(\bfr)
\nonumber \\
&+& Z_1 Z_2 \alpha \, v_\Lambda(r, 0).
\label{V1S0}\eea
The coefficients $C_{0,2}^{\rm 1S0}$ differ in the $pp$ channel
from those in the $np$ channel due to the
charge-independence breaking interactions.
\vskip 0.3cm

$\bullet$ {\bf The deuteron ${}^3S_1-{}^3D_1$ coupled channel}:
The potential is of the form
\bea
V(\bfr; {}^3S_1-{}^3D_1) &=&
V_C^d(r) + S_{12}(\hat r) \, V_T^d(r)
\eea
where, due to eq.(\ref{dd}),
\bea
V_C^d(r) &=&
- \alpha_\pi\, v_\Lambda(r, m_\pi)
+ \frac{4\pi}{M} \left[ C_0^{\rm 3S1} -
 \Lambda^2 (\Lambda^2 r^2 -3 )\, C_2^{\rm 3S1}
 \right] \delta^3_\Lambda(\bfr),
\nonumber \\
V_T^d(r) &=&
- \alpha_\pi\, t_\Lambda(r, m_\pi)
- \frac{4\pi}{M} \frac{D_2^{\rm 3S1}}{\sqrt 8}
 \, \Lambda^4 r^2\, \delta_\Lambda^3(\bfr) .
\label{V3S1}\eea

With the regulator (\ref{Sdef}), it is a simple matter
to work out the explicit forms of the quantities we need:
\bea
\nabla^2_\bfr \delta_\Lambda^3(\bfr)
 = \Lambda^2 (\Lambda^2 r^2 -3 )\, \delta_\Lambda^3(\bfr),
\ \ \
r\frac{\del}{\del r} \frac{1}{r} \frac{\del}{\del r}
 \delta_\Lambda^3(\bfr)
= \Lambda^4 r^2\, \delta_\Lambda^3(\bfr)
\label{dd}\eea
and
\bea
\delta_\Lambda^3(\bfr) &=&
 \frac{\Lambda^3 \exp\left[- \frac{\Lambda^2 r^2}{2}\right]}
{(2\pi)^{\frac32}},
\\
v_\Lambda(r, m) &=&
\frac{1}{2 r} \left[
 \e^{-m r} \, \erfc\left(
  \frac{- \Lambda r + \frac{m}{\Lambda}}{\sqrt 2} \right)
 - (r \rightarrow -r) \right],
\\
v_\Lambda(r, 0) &=&
\frac{1}{r} \,
 \erf\left( \frac{\Lambda r}{\sqrt 2} \right)
\eea
where $\erf(x)$ and $\erfc(x) \equiv 1 - \erf(x)$ are the error functions,
\be
\erf(x) \equiv \frac{2}{\sqrt \pi} \int_0^x \! dt \,\e^{- t^2} .
\ee

Finally the radial functions, $u_0(r)$ for $^1S_0$ and
$u_d(r)$ and $w_d(r)$ for
${}^3S_1-{}^3D_1$, are obtained by solving the Schr\"odinger equations
\be
u_0''(r) +  M \left[ E - V(r; {}^1S_0)\right] u_0(r) = 0,
\ee
and
\bea
u_d''(r) + M \left[E - V_C^d(r)\right] u_d(r) &=&
  \sqrt{8} M V_T^d(r) w_d(r),
\nonumber \\
w_d''(r)  - \frac{6}{r^2} w_d(r)
 + M \left[E - V_C^d(r) +  2 V_T^d(r)\right] w_d(r) &=&
  \sqrt{8} M V_T^d(r) u_d(r),
\eea
with the boundary condition
\be
u_0(0) = u_d(0) = w_d(0) = 0.
\ee

One caveat is in order here. Solving the Schr\"odinger equation
with the irreducible vertex calculated to NLO is consistent with
the Weinberg scheme. As pointed out in \cite{cohen,KSWPDS},
however, it is not strictly consistent with the chiral counting
order by order since the Gaussian
cut-off function brings in all powers of $p/\Lambda\sim Q/\Lambda$
and so does the iteration to all orders of the irreducible vertex
that is involved in the evaluation of reducible diagrams.
Nevertheless, we believe that our procedure is safe from
uncontrollable errors.
First of all, in our treatment, the cutoff has a physical meaning
and should not be sent to infinity. Therefore there is no problem
of the sort described by Beane {\it et al.}\cite{cohen} and by Luke
and Manohar~\cite{lm}, which one would encounter if one were to
send the cutoff to infinity or if one were to use the {\it naive}
dimensional regularization (DR) in the Weinberg scheme. Furthermore
to the order we work in computing the irreducible diagrams,
our method is consistent with the power counting. Possible
errors reside only in higher-order terms that are being dropped.
A similar conclusion was reached by Gegelia~\cite{gegelia}.

\section{Renormalization}
\indent

For a finite cutoff $\Lambda$, only finite quantities enter.
However renormalization is required to render the theory consistent
(see \cite{donoghue} for details). In particular, certain cutoff-dependent
terms have to be absorbed into the constants before
determining them from empirical data. We proceed as follows. For a
given cutoff and for each channel, we determine the constants $C$'s
and $D$'s in (\ref{Vq}) by relating them to the low-energy
scattering parameters appearing in the effective-range formula.
\vskip 0.3cm

$\bullet$ {\bf The $np$ $^1S_0$ channel}:
The effective-range formula for this channel is
\be
p \cot\delta_0 = -\frac{1}{a_0} + \frac12 r_0 p^2
 + \calO(p^4)
\ee
where $p\equiv \sqrt{ME}$ is the relative momentum and the
subscript `0' denotes the $np$ ${}^1S_0$ channel, i.e.,  $\delta_0
\equiv {}^1\delta_{np}$ and similarly for the scattering length $a_0$
and the effective range $r_0$. The low-energy scattering parameters, 
$a_0$ and $r_0$, can be given by the radial wavefunction 
$u_0(r)$ at zero energy:
\bea
a_0 &=& \left. \frac{r u_0'(r) - u_0(r)}{u_0'(r)}\right|_{r \ge R},
\label{a1S0}\\
r_0 &=& 2 \int_0^\infty \! dr \,
\left[\phi_0^2(r) - u_0^2(r) \right],
\label{r1S0}\eea
where $R$ denotes the range of the potential
such that V(r)=0 for $r\ge R$,
and $\phi_0(r)$ is the large-$r$ asymptotic
wavefunction of $u_0(r)$,
\be
\lim_{r\rightarrow \infty} u_0(r) = \phi_0(r)
 = 1 - \frac{r}{a_0}.
\ee
The experimental values of $a_0$ and $r_0$ which we shall
use to fix $C_0^{np}$ and $C_2^{np}$ are:
\be
a_0^{\rm exp}= -23.749(8)\ \fm,
\ \ \
r_0^{\rm exp}= 2.81(5)\ \fm.
\label{np:exp}\ee
\vskip 0.3cm

$\bullet$  {\bf The $pp$ ${}^1S_0$ channel}: In this channel,
the infinite-ranged
Coulomb potential must be taken into account. There are some
subtleties in doing this because there are many different
definitions of the low-energy scattering parameters depending on how the
long-ranged $\calO(\alpha^2)$ corrections are treated \cite{bcss}.
One reliable scheme was described in full detail in \cite{pkmr-apj},
which we will follow here.  We define the phase shift $\delta^C$
-- and consequently the
low-energy scattering parameters
$a^C$ and $r^C$ -- in
the {\it absence} of the potentials of order $\calO(\alpha^2)$ and take into
account these higher order EM effects by perturbation theory.
The expressions obtained in \cite{pkmr-apj} for the
low-energy scattering parameters are
\bea
a^C &=& \frac{1}{m_p \alpha} \,\left.
 \frac{u_{\rm C+N}'(r) \phi_2(r) - u_{\rm C+N}(r) \phi_2'(r)}
   {u_{\rm C+N}'(r) \phi_1(r) - u_{\rm C+N}(r) \phi_1'(r)}
 \right|_{r \ge R},
\label{aC}\\
r^C &=& 2 \int_0^\infty \! dr \,
\left[\phi_C^2(r) - u_{\rm C+N}^2(r) \right]
\label{rC}\eea
with
\be
\phi_C(r) = \phi_1(r) - \frac{\phi_2(r)}{m_p \alpha\, a^C},
\ee
where $m_p\simeq M$ is the proton mass, $u_{\rm C+N}(r)$ is the
radial wavefunction that approaches the function $\phi_C(r)$
asymptotically, and $\phi_1(r)$ and $\phi_2(r)$ are the radial
wavefunctions calculated with only the Coulomb interaction at zero
energy,
\bea
\phi_1(r) &=& 2 \sqrt{m_p \alpha r} K_1(2 \sqrt{m_p \alpha r}),
\nonumber \\
\phi_2(r) &=& \sqrt{m_p \alpha r} I_1(2 \sqrt{m_p \alpha r}).
\eea
Here $K_1(z)$ and $I_1(z)$ are the modified Bessel functions.
In fixing $C_0^{pp}$ and $C_2^{pp}$, we shall use the following
experimental values
(extracted from the Nijmegen multi-energy analysis \cite{nijmegen})
of $a^C$ and $r^C$:
\be
a^C_{\rm exp}= -7.8196(26)\ \fm,
\ \ \
r^C_{\rm exp}= 2.790(14)\ \fm.
\label{pp:exp}\ee
\vskip 0.3cm

$\bullet$  {\bf The deuteron $^3S_1-^3D_1$ channel}: In this channel,
 we have the choice of renormalizing the
theory either at zero energy or at the deuteron pole position
$E= -B_d$, where $B_d$ is the deuteron binding energy.
Here we shall adopt the latter.
The coefficient $D_2^d$ will be determined by
the experimentally well-known deuteron $D/S$ ratio
$\eta_d$.
Near the deuteron pole position, the effective-range
formula has the form
\be
p \cot\delta_d
= - \gamma + \frac12 \rho_d (p^2+\gamma^2)
 + \calO\left[(p^2+\gamma^2)^2\right]
\ee
where $\delta_d$ is the $S$-wave eigenphase and
$\gamma\equiv \sqrt{M B_d}$
is the deuteron wave number.
Thus the quantities $B_d$, $\rho_d$ and $\eta_d$ are  inputs to
determine the three constants $C_0^d$, $C_2^d$ and $D_2^d$.
Since by construction our potential
is energy-independent,
the wavefunction normalization factor $A_s$ is not an independent
quantity. It is related to the input parameters by
\be
A_s^2 = \frac{2 \gamma}{(1 + \eta_d^2) (1 - \gamma \rho_d)} .
\label{AsER}\ee
In our numerical calculation, we shall use
the following experimental values for $B_d$, $A_s$ and $\eta_d$:
\be
B_d^{\rm exp} = 2.224575(9)\ \MeV, \ \
A_s^{\rm exp} = 0.8846(8)\ \fm^{\frac12}, \ \
\eta_d^{\rm exp} = 0.0256(4).
\label{d:exp}\ee
{}From eq.(\ref{AsER}) we get $\rho_d
= 1.7635(46)\ \mbox{fm}$.

\section{Results}
\indent

As stated, we shall use the effective-range parameters to fix
the coefficients of (\ref{Vq}) for the $np$ and $pp$  $^1S_0$
channels and $B_d$, $A_s$ and $\eta_d$ for the deuteron channel.
Given solutions for the Schr\"odinger equation
with the nucleon-nucleon potential thus obtained
and
the electroweak current
derived
to the leading order, we
are in a position to carry out
a {\it parameter-free} calculation of the M1 matrix element for the $np$
capture process and the Gamow-Teller matrix element for the $pp$
fusion process.
Furthermore,
the deuteron static properties such as the charge
radius $r_d$, the quadrupole moment $Q_d$ and the magnetic moment
$\mu_d$ can be predicted.
 The precise value of the cutoff is of
course not determined by the theory
but the consistency of the scheme requires that
the results should be insensitive to its precise value provided
that it is picked judiciously: e.g., around the pion mass if the pion is
integrated out and around the two-pion mass if the pion is present.

The relevant electroweak matrix elements we shall focus on are
\footnote{We define these matrix elements with all other factors stripped
off. In \cite{pkmr-prc} and \cite{pkmr-apj},
$\calM_{\rm M1}$ and $\calM_{\rm GT}$
were denoted
by $M_{\rm 1B}$ and $M_{\rm 1B}^{\rm C+N}$, respectively.}
\bea
\calM_{\rm M1} &\equiv& \int_0^\infty\! dr\, u_d(r) u_0(r),
\\
\calM_{\rm GT} &\equiv& \int_0^\infty\! dr\, u_d(r) u_{\rm C+N}(r).
\eea
The ${\cal O}(\alpha^2)$ (VP and C2) corrections
to the GT matrix element
are calculated perturbatively\cite{pkmr-apj}
using the potentials
\bea
V_{\rm VP}(r) &=&
 \frac{2\alpha^2}{3\pi} \int_1^\infty\! dx\,
 \e^{-2 m_e r x} \left(1 + \frac{1}{2 x^2} \right)
 \frac{\sqrt{x^2-1}}{x^2},
\nonumber \\
V_{\rm C2}^{\rm C}(r) + V_{\rm C2}^{\rm N}(r)
 &=& - \frac{\alpha}{m_p r} \left(\frac{\alpha}{r} + V_N(r)\right),
\eea
where $m_e$ is the electron mass,
and $V_N(r)$ denotes the potential (\ref{V1S0})
but without the Coulomb interaction.

\subsection{Without the pion}
\indent

We first discuss the case where the one-pion-exchange term in
eq.(\ref{Vq}) is absent.
The results with the NLO terms are given in Table \ref{table:2}.
There the scattering lengths and effective ranges 
for the $np$ and $pp$ channels and $B_d$, $A_s$ and $\eta_d$ 
for the deuteron channel are used
as inputs to determine the constants of (\ref{Vq}).
We see that the results as a whole are stable
against changes in the cutoff as long as the cutoff is
taken above the pion mass.
The theoretical values are compared with experimental data when available
and with the Argonne $v_{18}$ results\footnote{
As we discuss below, the results given by the Argonne $v_{18}$ wavefunctions
are fit to experiments and hence can be considered as ``experiments".}
otherwise.
There is good agreement between the calculated
and experimental values.
\begin{table}[tbp]
\caption{\protect \small
The next-to-leading order (NLO) results
{\em without} the pion field.
The static properties of the deuteron,
the M1 and GT amplitudes and
the VP and C2 contributions to the GT amplitude are listed
for various choices of
the cutoff $\Lambda$.
The low-energy input parameters are the
scattering lengths and effective ranges for the $np$ and $pp$
${}^1S_0$ channels,
and $B_d$, $A_s$ and $\eta_d$ for the deuteron channel.
Their experimental values are given by
eqs.(\ref{np:exp}, \ref{pp:exp}, \ref{d:exp}).}
\label{table:2}
\begin{center}
\begin{tabular}{|c|cccccc|cc|} \hline
$\Lambda$ (MeV) & $140$ & $150$ & $175$ & $200$ & $225$ & $250$ &
 Exp. & $v_{18}$\cite{v18} \\
\hline \hline
$r_d$ (fm)
   & 1.973 & 1.972 & 1.974 & 1.978 & 1.983 & 1.987 & 1.966(7) & 1.967 \\
$Q_d$ ($\mbox{fm}^2$)
  & 0.259 & 0.268 & 0.287 & 0.302 & 0.312 & 0.319 & 0.286 & 0.270 \\
$P_D$ (\%)
  & 2.32  & 2.83 & 4.34 & 6.14 & 8.09 & 9.90 & $-$ & 5.76 \\
$\mu_d$
  & 0.867 & 0.864 & 0.855  & 0.845 & 0.834 & 0.823  & 0.8574 & 0.847 \\
\hline
$\calM_{\rm M1}$ (fm)
   & 3.995 & 3.989 & 3.973 & 3.955 & 3.936 & 3.918 & $-$ & 3.979 \\
\hline
$\calM_{\rm GT}$ (fm)
  &4.887 & 4.881 & 4.864 & 4.846 & 4.827 & 4.810 & $-$ & 4.859 \\
$\delta_{\rm 1B}^{\rm VP}$ (\%)
 & $-0.45$ & $-0.45$ & $-0.45$ & $-0.45$ & $-0.45$ & $-0.44$ & $-$ & $-0.45$\\
$\delta_{\rm 1B}^{\rm C2:C}$ (\%)
 &0.03 & 0.03 & 0.03 & 0.03 & 0.03& 0.03 & $-$ & 0.03 \\
$\delta_{\rm 1B}^{\rm C2:N}$ (\%)
 & $-0.19$ & $-0.19$ & $-0.18$ & $-0.15$ & $-0.12$ & $-0.10$ & $-$ & $-0.21$ \\
\hline
\end{tabular} \end{center} \end{table}

\begin{figure}[htbp]
\centerline{\protect \epsfig{file=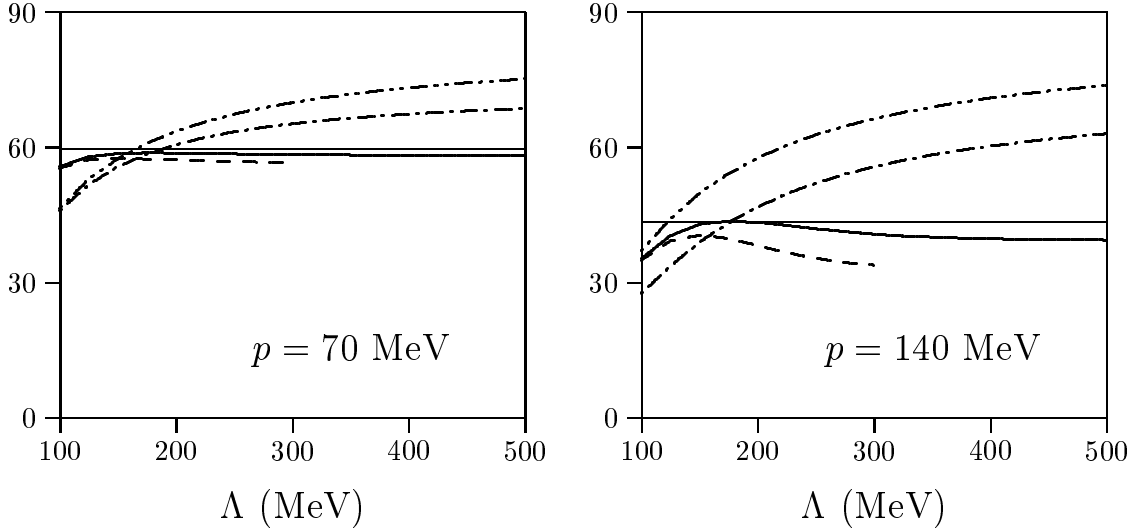}}
\caption[deviate]{\protect \small
The $np$ $^1 S_0$ phase shift (degrees) vs.
the cutoff $\Lambda$ for fixed CM momenta
$p=70$ MeV (left) and $p= 140$ MeV (right).
The next-to-leading order (NLO) results are given by the
solid (with pions) and dashed (without pions) curves,
and the leading-order (LO) results by the
dot-dashed (with pion) and dot-dot-dashed (without pion) curves.
The horizontal line represents the experimental data
obtained from the Nijmegen multi-energy analysis.
The NLO result without
pions is drawn only up to $\Lambda \gsim 300$ MeV, above which
the theory becomes meaningless, that is, unnatural as well as
inconsistent.\label{figure1}}
\end{figure}

These results incorporate
up to NLO terms.
Now, what happens when one limits the
calculation to the leading order?
The answer is that it does not work.
It is not difficult to understand why this is so.
Doing a leading order (LO) calculation corresponds to ignoring
the $q^2$ term in (\ref{Vq})
and, without the $q^2$ term -- and without the pion,
there will be no way to reproduce all the deuteron properties
as well as the electroweak matrix elements.
As shown in Fig. \ref{figure1} for fixed CM momenta
$p\sim m_\pi/2$ and $p\sim m_\pi$,
the low-energy $np$ $^1S_0$ phase shifts calculated in LO
(denoted by the dot-dot-dashed curve in Fig. \ref{figure1})
depend strongly on the cutoff.
This strong cutoff dependence,
however,
is considerably reduced,
when the NLO term is included (dashed curve).
The improvement is
more dramatic, the lower the momentum probed is.
The introduction of the pion field makes the
range of the relevant cutoff considerably greater,
approaching the next momentum scale involved,
namely, $\gsim 2m_\pi$. Thus the consistency
condition for a viable effective theory with a given set of effective
degrees of freedom requires going to
sufficiently high order
--  NLO in the present case. To what order the calculation needs to
be pushed and what additional degrees of freedom are required
will of course depend on the problem one is addressing.

\subsection{With the pion}
\indent

We shall consider the cases with and without the ${\cal O}
(q^2)$ term in (\ref{Vq}), the former corresponding to LO and the
latter to NLO.
\vskip 0.3cm

$\bullet$ {\bf The LO calculation}: Here the scattering lengths
for the $^1S_0$ channel and the deuteron binding energy $B_d$
are inputs. The resulting predictions are given in Table
\ref{table:pi0}.
\begin{table}[tbp]
\caption{\protect \small
The leading order (LO) results {\it with} pions. Note that
only the scattering lengths for the ${}^1S_0$ channel and the
deuteron binding energy are inputs. 
All the quantities except $C_0$'s are predicted;
in particular, the wavefunction normalization $A_s$ and 
the $D/S$ ratio $\eta_d$ are {\it not} input but predicted
here.}
\label{table:pi0}
\begin{center}
\begin{tabular}{|c|ccccc|cc|} \hline
$\Lambda$ (MeV) & $300$ & $400$ & $500$ & $600$ & $700$ &
 Exp. & $v_{18}$\cite{v18} \\
\hline \hline
$C_0^{np}\ (\fm)$ & $-0.828$ & $-0.621$ & $-0.501$ &
  $-0.422$ & $-0.366$ &  & \\
$C_0^{pp}\ (\fm)$ & $-0.810$ & $-0.615$ & $-0.500$ &
  $-0.423$ & $-0.367$ &  &  \\
$C_0^{d}\ (\fm)$ & $-0.979$ & $-0.384$ & $-0.144$ &
  $0.795$ & $0.789$ &  &  \\
\hline
$A_s\ (\fm^{-\frac12})$ &
  0.853 & 0.854 & 0.861 & 0.866 & 0.869 & 0.8846(8) & 0.885 \\
$\eta_d$ &
 0.0227 & 0.0250 & 0.0260 & 0.0263 & 0.0265 & 0.0256(4) & 0.0250 \\
$r_d$ (fm)
   & 1.897 & 1.902 & 1.918 & 1.930 & 1.936 & 1.966(7) & 1.967 \\
$Q_d$ ($\mbox{fm}^2$)
   & 0.218 & 0.249 & 0.266 & 0.274 & 0.277 & 0.286 & 0.270 \\
$P_D$ (\%)
   & 2.65 & 4.59 & 5.97 & 6.73 & 7.08 & $-$ & 5.76 \\
$\mu_d$
   & 0.865 & 0.854  & 0.846 & 0.841 & 0.839  & 0.8574 & 0.847 \\
\hline
$\calM_{\rm M1}$ (fm)
   & 4.176 & 4.186 & 4.164 & 4.140 & 4.122 & $-$ & 3.979 \\
$\calM_{\rm GT}$ (fm)
   & 5.054 & 5.063 & 5.042  & 5.019 & 5.002 & $-$ & 4.859 \\
\hline
\end{tabular}
\end{center}
\end{table}
With the pion, the deuteron does exist with its properties
qualitatively reproduced. This is of course not surprising
to nuclear physicists since the
significant
role of the pion in the deuteron structure has been
known for a long time.
What is surprising, however,
is that even the electroweak matrix elements
are not far from the ``experimental values",
though some $4\sim 5\ \%$ higher.
The effective ranges are not controlled without the NLO potential,
so it is clear that the phase shifts can be reproduced
only for very low energies.
Actually we observe in Fig. \ref{figure1} that
while the LO result with the pion (denoted by dot-dashed)
is better than the LO results without the pion
(dot-dot-dashed curve),
it is much worse than the NLO result without the pion
(dashed curve).
This feature, visible also in Fig. \ref{figure2},
will be discussed in more detail later.
An important point to note here is that
when the pion field is present,
the cutoff climbs up to roughly twice the pion mass.
This is consistent with our expectation.
\vskip 0.3cm

$\bullet$ {\bf The NLO calculation}: The pion contribution is fixed
by chiral symmetry, so the parameters of the theory are determined
in the same way as for
the NLO calculation in the absence of the pion. The results are
given in Table \ref{table:pi2}.

The presence of the pion in the NLO calculation brings a noticeable
improvement in the accuracy of the prediction making the
dependence on the cutoff markedly
weaker.
This is true not only
for
the static properties (Table \ref{table:pi2}) but
also
for
the phase shift
for $p \approx m_\pi$
(see the right panel of Fig.\ref{figure1}).
The cutoff required is roughly twice that of the pionless case
and this again is consistent with our expectation.
Note that the charge radius,  the quadrupole moment
and the magnetic moment of the deuteron all come out
very close to the experimental values,
indicating that exchange-current corrections,
if any, should be tiny.
As remarked in the discussion section,
this is expected on a general ground.
The M1 matrix element and
the GT matrix element receive a few per cents
exchange-current corrections
which are well controlled for the former and somewhat less well
controlled for the latter. See later on this matter.

\begin{table}[tbp]
\caption{\small
The NLO results {\em with} pion field.
See the caption of Table \ref{table:2}.}
\label{table:pi2}
\begin{center}
\begin{tabular}{|c|cccccc|cc|} \hline
$\Lambda$ (MeV) & $200$ & 250 & $300$ & 350 & $400$ & $500$ &
 Exp. & $v_{18}$\cite{v18} \\
\hline \hline
$r_d$ (fm)
   & 1.963 & 1.965 & 1.966 & 1.967 & 1.968 & 1.968 & 1.966(7) & 1.967 \\
$Q_d$ ($\mbox{fm}^2$)
   & 0.261 & 0.268 & 0.272 & 0.273 & 0.274 & 0.274 & 0.286 & 0.270 \\
$P_D$ (\%)
   & 3.16 & 4.11 & 4.77 & 5.16 & 5.35 & 5.39 & $-$ & 5.76 \\
$\mu_d$
   & 0.862 & 0.856 & 0.853 & 0.850 & 0.849 & 0.849 & 0.857 & 0.847 \\
\hline
$\calM_{\rm M1}$ (fm)
   & 3.987 & 3.976 & 3.968 & 3.963 & 3.958 & 3.952 & $-$ & 3.979 \\
\hline
$\calM_{\rm GT}$ (fm)
 & 4.884 & 4.874 & 4.867 & 4.862 & 4.859 & 4.854 & $-$ & 4.859 \\
$\delta_{\rm 1B}^{\rm VP}$ (\%)
 & $-0.46$ & $-0.45$ & $-0.45$ & $-0.45$ & $-0.45$ & $-0.45$ & $-$ & $-0.45$\\
$\delta_{\rm 1B}^{\rm C2:C}$ (\%)
 &0.04 &0.03 & 0.03 & 0.03 & 0.03 & 0.03 & $-$ & 0.03 \\
$\delta_{\rm 1B}^{\rm C2:N}$ (\%)
 & $-0.24$ & $-0.20$ & $-0.15$ & $-0.14$ & $-0.21$ & $-0.42$ & $-$ & $-0.21$ \\
\hline
\end{tabular} \end{center} \end{table}

\begin{figure}[hbtp]
\centerline{\protect \epsfig{file=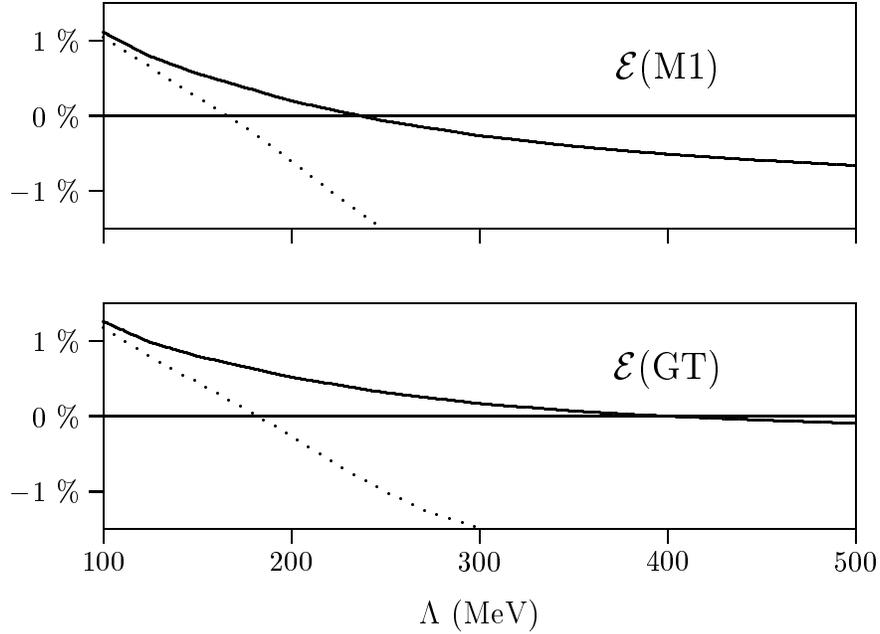}}
\caption{\protect \small
${\cal E}(\mbox{M1})$ (upper) and
${\cal E}(\mbox{GT})$ (lower) vs. the cutoff $\Lambda$.
The solid curves represents the NLO results with
pions and the dotted curves without pions.
\label{E-MGT}}
\end{figure}

To give a more quantitative idea
as to how much the incorporation of the pionic degree of freedom
reduces the cut-off dependence of our calculation,
it is informative to consider
${\cal E}(\mbox{M1})$ and ${\cal E}(\mbox{GT})$
defined by
\be
{\cal E}(\mbox{M1}) \equiv \frac{\calM_{\rm M1}^{\rm th}
 - \calM_{\rm M1}^{v18}}{\calM_{\rm M1}^{v18}},
\ \ \
{\cal E}(\mbox{GT}) \equiv \frac{\calM_{\rm GT}^{\rm th}
 - \calM_{\rm GT}^{v18}}{\calM_{\rm GT}^{v18}},
\ee
where $\calM_{\rm M1}^{\rm th}$
and $\calM_{\rm M1}^{v18}$ denote,
respectively,
the M1 transition matrix element of our NLO calculation
and that of the Argonne potential,
and similarly for ${\cal E}(\mbox{GT})$.
Here we are taking the Argonne potential as ``experiment"
(see the next subsection for more explanation on this point).
In Fig. \ref{E-MGT} are plotted the deviations from
the ``experiment" as a function of the cutoff.
We see that as in the case of the phase shift for $p\sim m_\pi$,
the inclusion of the pion degree of freedom markedly
reduces the cutoff dependence,
in conformity with the requirement
for a viable effective theory
that the more accurate the power expansion,
the less cutoff-dependent physical observables
should become.

\subsection{Comparison with a realistic phenomenological potential}
\indent

In our earlier calculations of the $np$ capture process \cite{pmr}
and the proton fusion process \cite{pkmr-apj},
we resorted to the wavefunctions obtained from the
Argonne $v_{18}$ potential -- which is
one of the most realistic phenomenological potentials fit to {\it
all} available
two-nucleon data up to $T_{\rm Lab}$=350 MeV --
to calculate the matrix elements
of the M1 and GT operators derived to the
next-to-next-to leading order (NNLO) in chiral expansion.
It might appear to some purists that this ``hybrid" approach
bringing in information from outside
of the chiral perturbation theory framework is not consistent
with the premise of an effective theory
and hence throws doubt on the value and validity
of such calculations.
In this subsection, we argue that such a doubt
can be dispelled because the hybrid model
turns out to be in precise numerical agreement
with the effective theory scheme.
For our purpose, it suffices to limit our consideration to the
leading-order matrix elements.

In \cite{pkmr-prc},
we showed that the leading-order M1 matrix element
calculated in the hybrid method was
is in good agreement with the result of
the effective field theory approach
in the absence of the pion.
Here we confirm that the M1 matrix element
as well as the GT matrix element (figuring in the $pp$ fusion process) 
calculated {\it both with and without} pions
quantitatively support the equivalence of the hybrid approach
and the effective field theory approach with cutoff.
We see from the NLO calculations given in Tables \ref{table:2} and
\ref{table:pi2} that the electroweak matrix elements calculated in
NLO in the effective field theory agree with those computed with
the Argonne $v_{18}$ within 1 \% error,
with very little dependence on the cutoff.
With the pion included the agreement is even more
striking as one can see in Figure \ref{E-MGT}.

Comparison is made also for the ${\cal O} (\alpha^2)$ EM
corrections to the $pp$ fusion matrix element. These are obviously
small corrections, smaller than higher-order power corrections
ignored in the calculation but they cannot be ignored in the solar
proton burning process as discussed in \cite{bahcall}. As one can
see in Tables \ref{table:2} and \ref{table:pi2}, the agreement is
again almost perfect.

\section{Meson-Exchange Currents and Short-Range Correlation}
\indent

So far we have considered the matrix elements of only one-body
electroweak currents.
In considering meson-exchange currents
that appear as higher-order corrections,
it is useful to recall the ``chiral filter mechanism" first
studied in \cite{rho91,KDR,pmraxial}.
The chiral filter
mechanism is the statement that,
whenever soft-pion exchange is {\it not forbidden}
by symmetry or {\it kinematically unsuppressed},
the dominant term is given by chiral symmetry
and the corrections to it should be suppressed.
A corollary to this is that if soft-pion exchange is
{\it forbidden} by symmetry or {\it kinematically suppressed},
then higher-order terms enter non-negligibly
so that the predictive power of the standard chiral perturbation
strategy is appreciably weakened if not lost.

According to this argument, we can make the following qualitative
statements:
\begin{itemize}
\item The corrections to the single-particle M1 matrix
element in the $np$ capture process\footnote{The same discussion
applies to the time component of the axial current, that is, the
axial-charge transitions (first-forbidden $\beta$-decays) in nuclei.
For detailed discussions, see \cite{pmraxial}.}
are given by one-soft-pion exchange (NLO) and loop corrections (NNLO).
Since the LO one-body operator comes from the vertex
with $\bar \nu_i=1$, these corrections are, respectively,
$\calO(Q)$ and $\calO(Q^3)$ relative  to the LO operator.
Working to  $\calO(Q^3)$ relative to
the (non-vanishing) leading order,
we need to have wavefunctions (or potentials)
 accurate up to $\calO(Q^3)$ for the
one-body matrix element, to $\calO(Q^2)$ for the
soft-pion exchange and to $\calO(Q^0)$ for one-loop
correction. Since parity allows only even powers
in the potential, it is sufficient therefore to have
an $\calO(Q^2)$ accuracy.  Our effective field theory to NLO
meets this requirement.
Let $\calM_{\rm M1}^{\rm total}= \calM_{\rm M1} (1 + {\cal R})$
be the matrix element of the full M1 operator,
with ${\cal R}$ being the ratio
of the matrix element of the exchange current operator
to that of the leading M1 operator, ${\cal M}_{\rm M1}$.
With a ``low-quality" phenomenological potential,
we will very likely get a poor estimate
of the leading one-body matrix element
$\calM_{\rm M1}$
mostly because of its inability to reproduce
the low-energy scattering parameters.
Even so, as explained above, the ratio ${\cal R}$ calculated
with the same wavefunction
is expected to be much less sensitive to the quality
of the phenomenological potentials.
That is, ${\cal R}$ should be
to a large extent
model-independent.
Another way of arriving at this
conclusion
is as follows.
As we have shown in this article,
the one-body matrix element $\calM_{\rm M1}$ can
be calculated with great accuracy
with the NLO EFT potential as well as with a
``high-quality" phenomenological potential
such as the Argonne $v_{18}$,
both of which reproduce all the relevant low-energy
scattering observables with high precision.
Since the total matrix element $\calM_{\rm M1}^{\rm total}$
should be the same regardless of
whether we adopt the EFT framework developed here
or the Argonne $v_{18}$ wavefunctions,
we conclude that the ratio ${\cal R}$ should not
depend on which of the two methods is used.

One non-trivial consequence of this argument is
that the ratio ${\cal R}$ should not depend upon
how the current is regularized.
Since the NLO correction term is made
of tree diagrams only,
the ratio should be reliably
given by the Argonne $v_{18}$ wavefunction.
As for the NNLO term, it was found in \cite{pmr}
that all but one term (hereafter referred to as $\kappa$ term) is
fixed by experiments.
This $\kappa$ term which is left undetermined  by experiment
arises from a NNLO two-body counter term in the current.
If one were to use the hybrid method of \cite{pmr},
this $\kappa$ term -- which is of zero range -- could not contribute
since it is suppressed by the
short-range correlation in the wavefunction. Furthermore all zero-range
operators that would appear as higher-order counter terms would also
be suppressed. On the other hand, if
one were to {\it naively} calculate the matrix element of the same
zero-range operators in an EFT with the cut-off regularization used
in this paper, the contribution would be non-zero unless the constants
(e.g., $\kappa$) happen to be all zero. So it would seem
that the two approaches give different results for non-zero
$\kappa$'s at NNLO and higher orders.
We conjecture that this difference can be at least partially resolved
by appealing to the operator product expansion (OPE)
in the wavefunction discussed in the context of EFT by
Lepage~\cite{lepage}.
The OPE implements in the wavefunction the physics of
perturbative QCD which is left out in the EFT framework
discussed here.
For long-wavelength operators (such as the single-particle
M1 and GT operators), the effect of OPE would not be important. 
However, operators probing short-distance physics
such as the zero-range $\kappa$ term
should be affected by the OPE factor in the wavefunction.
Our conjecture is that the short-range correlation
in the standard nuclear physics approach and the OPE
in the EFT approach a la Lepage capture qualitatively the same physics.
At present we do not know how to make the connection between
the two descriptions. We suggest that the lack of
knowledge on the precise connection accounts for the uncertainty in the former
approach manifested in different results for different forms of short-range
correlation functions and for different short-range correlation sizes. This
uncertainty -- small in some cases like the $np$ capture -- is the price
to pay in ``killing" the higher-order counter terms, which we expect to be
unimportant for very low-energy processes we have been considering but
significant for higher-energy probes.

For completeness we mention that an NNLO calculation for the
exchange current contribution performed using the Argonne $v_{18}$
wavefunctions gave $\sim 10 \%$ contribution to the $np$ capture
cross section leading to a precise agreement with the experiment~\cite{pmr}.
The major contribution was given by the dominant soft-pion term, with
only about 1 \% coming from NNLO loop corrections. The uncertainty
in the calculation was about 1 \% in the cross section, representing
the above-mentioned
uncertainty in the short-range correlation that was used to ``kill"
the undetermined delta-function counter term.
\item
The power counting of the GT matrix element is slightly
different from that of the M1 operator.
Here the one-body GT operator is not suppressed
by symmetry
but one-soft-pion-exchange current is.
As a consequence, the leading correction (one-pion-exchange)
is of order of $\calO(Q^3)$ while loops enter at
$\calO(Q^4)$ or higher compared to the one-body operator.
This means that the tree-order
contributions involving one pion-exchange two-body operators
and  loop contributions and/or
higher-dimensional counter terms that come at next order must
be comparable in importance. So far a consistent calculation
exists only in tree order, which predicts a $\sim 4 \%$ correction
(an enhancement) to the leading GT matrix
element squared\cite{pkmr-apj}.
The (corollary to the) chiral filter mechanism implies
that this correction could receive
a substantial contribution from higher-order or short-distance
corrections. We should mention that the Landau-Migdal correction to
$g_A$ in medium discovered a long time ago\cite{gA*}\footnote{
\protect A more modern interpretation in heavy-baryon chiral perturbation
theory is given in \cite{pjm}.}
corresponds to this class of effects.
\item The exchange-current corrections to the deuteron properties, e.g.,
the charge radius, the quadrupole moment and the magnetic moment, are
also suppressed by the chiral filter \cite{rho91,KDR}.
An additional suppression
occurs
due to isospin symmetry.
This explains why in Tables \ref{table:2} and \ref{table:pi2}
the theory without exchange current corrections agrees --
almost precisely --
with experiments without any further corrections.

\end{itemize}

\section{Breakdown of Effective Field Theory?}
\indent

\begin{figure}[tbh]
\centerline{\protect \epsfig{file=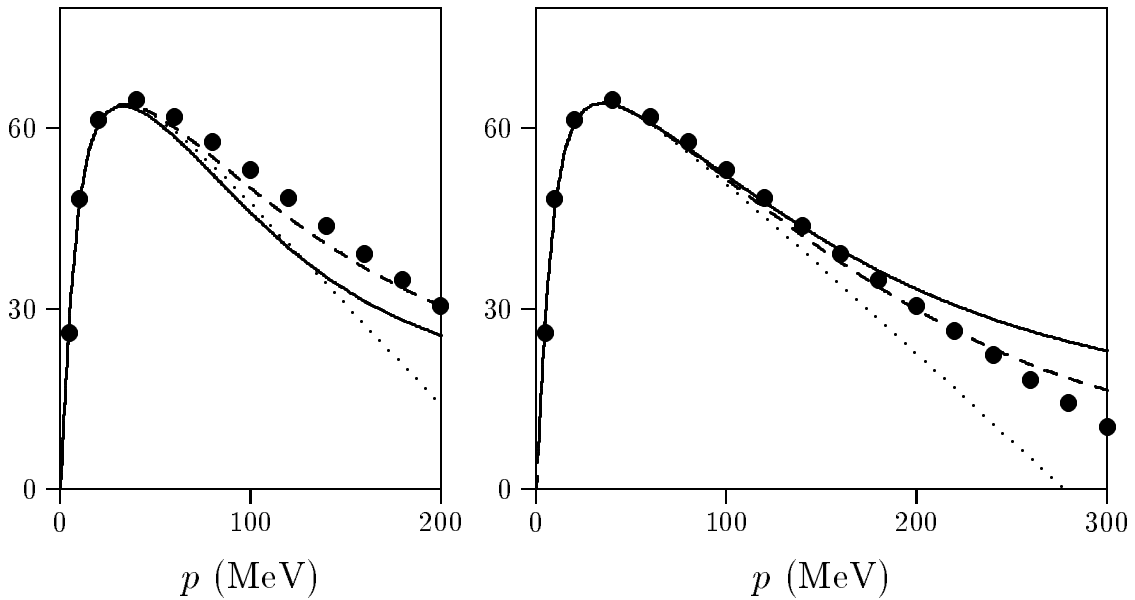}}
\caption[phase]{\protect \small
The $np$ $^1S_0$ phase shift (degrees) vs. the center-of-mass (CM)
momentum $p$. We show the predicted results both
without (left) and with (right) the pion field.
For each case, our NLO results are given
for two extreme values of the cutoff:
The solid curve represents the lower limit,
$\Lambda= 100$ MeV (without the pion) and $200$ MeV
(with the pion),
and the dotted curve the upper limit,
$\Lambda= 300$ MeV (without the pion) and $500$ MeV (with the pion).
The ``experimental points" obtained from the Nijmegen multi-energy analysis
are given by the solid circles.
Our theory with $\Lambda=150$ MeV (without the pion) and $\Lambda=
250$ (with the pion) are shown by the dashed line to show that the
theory is in almost perfect agreement with the data up to $p \lsim
200$ MeV. }\label{figure2}
\end{figure}

Given a cutoff and a set of
degrees of freedom chosen to work with, where is the given theory
expected to break down
and
become inaccurate?
By way of answering this question, let us look at
low-energy $np$ $^1S_0$ phase-shifts. In Fig. \ref{figure2} are
plotted the phase shifts vs. the CM momentum $p$
for the lower and upper limits of the cutoff
both in the absence and in the presence of
the pion degree of freedom.
The range $\Lambda=(100\sim 300)$ MeV represents the
range of the cutoff in which the NLO theory without pions
should be valid
and the range $\Lambda=(200\sim 500)$ MeV
is the corresponding range when the pion field is incorporated.

There are several lessons one can learn from Fig. \ref{figure2}.
By fiat, all the constants of the theory are
fixed by the scattering length and the effective range,
that is, the first two terms in the effective-range expansion
are fixed.
However the Gaussian cutoff introduces terms of
${\cal O}[(p/\Lambda)^n]$ for $n \ge 3$.
The theory works fairly independently of the cutoff
up to $p\sim 100$ MeV if the pion field is included,
and up to $p\sim 50$ MeV if the pion is absent;
beyond these limits
the cutoff dependence
becomes noticeable for both cases.
Since the two-term effective-range formula fits
data up to $p \sim 140$ MeV,\footnote{
This accuracy of the effective-range formula to $p\sim 140$ MeV
is somewhat unnatural from the point of view of EFT. It is perhaps fortuitous.
Intuitively one would expect that the effective-range formula
should work only up to $p \sim m_\pi/2$, since
the $S$-matrix element contains the left-hand cut starting at 
$p^2 = -(m_\pi/2)^2$.}
higher-order terms must be responsible for the
{\it deviation} from the experiment.
Thus, in order to recover the fit,
terms of order ${\cal O}[(p/\Lambda)^n]$ with $n \ge 3$
must be included in the (irreducible) potential $V$.
In fact the two-term effective-range formula works
so well due to the fact that
the third term of the effective-range formula,
effective volume term,
is very close to zero\cite{eff_vol}.
It then means that the two-term effective-range formula
is essentially the same with the three-term effective-range formula
with the effective volume term fine-tuned to zero.
In order for this to happen in our theory,
the contribution from the Gaussian cutoff and the
corresponding higher-order counter term
-- which must be added  --
should cancel to give a vanishingly small  volume term. Therefore
it would be ``unnatural" 
if our NLO calculation, without involving a ``fine-tuning",
were to work as well as the ``three-term" effective-range formula.
The left panel of Fig. 3 (NLO without the pion field)
is thus ``natural". On the other hand,
the right panel of Fig. 3 (NLO with the pion field)
has the accuracy comparable to the ``three-term"
effective-range formula, which should be understood
as the role of the pion field.

Given that we are truncating the series, thereby making the theory
manifestly approximate, it seems reasonable to exploit the
insensitivity to the precise value of the cutoff and do a ``best
fit." For this, we shall pick -- within the range allowed -- a specific 
value for cutoff in fitting experimental
data. We should stress that this has nothing to do with a
``fine-tuning" usually associated with an unnatural theory.
{}From Table \ref{table:pi2} we can see that, when the pion field is present,
$\Lambda\simeq 250$ MeV
gives an overall fit of the deuteron properties
as well as electroweak matrix elements.
In Fig. \ref{figure2} is plotted the phase shift prediction for this
cutoff. The agreement with experiments is excellent up to
$p\lsim 200$ MeV.\footnote{\protect
In \cite{KSWPDS}, an even more impressive agreement for the phase shift
was obtained with the pion field included. However it should be noted that
to achieve a better global fitting, the authors used values of the
coefficients of the potential that are different from the ones 
determined by their renormalization procedure. As a consequence,
the description for the threshold region became less reliable. 
}
Thus for this value of the cutoff, {\it all} low-energy
two-nucleon properties are accurately described by the theory.

We can do a similar ``best-fitting" in the absence of the pion field
and get a result which,
though not as good as in the presence of the pion field,
is still quite satisfactory as one can see in the left panel of
Fig. \ref{figure2} for $\Lambda\simeq 150$ MeV.

\section{Discussion}
\indent

We have shown in this paper that the idea of effective field theory
in its simplest form works remarkably well in
nuclear physics of two-nucleon systems at low energy.
We have used the cut-off regularization which renders the notion
of effective theory physically transparent.
It has been seen
that the cut-off regularization
also renders moot~\cite{gegelia2} the issue of
whether there is a breakdown in Weinberg's counting rule,
which consists in applying the chiral counting only to irreducible diagrams,
 and subsequently summing up an infinite series of reducible graphs
containing irreducible vertices of a specified order.

We have further shown that the ``hybrid approach'' used in
\cite{pmr} to
calculate the $n+p\rightarrow d+\gamma$ cross section
in the framework of chiral perturbation theory (in the Weinberg version)
is fully justified by a systematic effective field theory technique
with a finite cutoff.

One of the important by-products of the effective field theory
for nuclei is precise calculations of nuclear weak processes
that figure importantly in astrophysics.
In \cite{pkmr-apj}, we computed in the ``hybrid approach''
the
cross sections
for the solar proton burning process
$p+p\rightarrow d+e^++\nu_e$ to an accuracy of 1 \%,\footnote{
\protect The main caveat in \cite{pkmr-apj} is in the
meson-exchange contribution
which comes out to be about 4 \% when calculated to the
chiral order ${\cal O}(Q^3)$.
The higher order contribution starts from ${\cal O}(Q^4)$,
and may not be necessarily negligible compared with the
${\cal O}(Q^3)$ contribution.
So a more ``conservative" error estimate may be 4 \%,
instead of 1 \%.}
and this has provided strong support
for the estimate used by Bahcall and collaborators
for the solar neutrino problem\cite{bahcall,bahcall2,solarRMP}.
In particular it has resolved the possible controversy
as to whether
or not
a field theoretic approach would
give a result basically different
from the classic nuclear physics approach
which is based on
Schr\"odinger equation
with realistic phenomenological N-N interactions\cite{bahcall}.
Another potentially important by-product is that with
the pion field incorporated, we should be able to
treat meson-exchange currents to one-loop
order (that is, $\calO(Q^4)$
relative to the leading GT operator).
This would allow a model-independent calculation of
processes like\footnote{
We would like to thank John Bahcall for calling our attention
to this problem.} \cite{hep}
\be
{}^3\mbox{He} + p \rightarrow {}^4\mbox{He} + e^+ + \nu_e\label{hep}
\ee
which may be measured at the SuperKamiokande and SNO.

As often heralded by the practitioners
of the dimensional regularization (DR),
one of the advantages of DR
over
other regularizations
is that the DR scheme preserves chiral invariance
and Lorentz invariance.
Indeed,
while giving an excellent result in a non-relativistic regime,
the cut-off regularization as used here loses
manifest chiral invariance and Lorentz invariance.
This might make higher-order
loop calculations difficult, if not impossible.
Fortunately,
in the tree-order treatment of meson exchange currents
for the electroweak processes as well as the other two-nucleon
processes considered in this paper,
we did not have to deal with
chirally non-invariant counter terms.
However, this problem does show up
at the level of loop calculations.
We have not investigated this problem in detail
but in calculating the process in (\ref{hep}),
where meson-exchange current corrections
are expected to be more important than
in the cases we have considered so far,
this issue will have to be resolved.
A related issue was investigated by the authors of \cite{donoghue}, 
who suggest that the difficulty is not insurmountable.
Their work indicates a promising way
of dealing with loop corrections in a finite cut-off field theory
for the process (\ref{hep}) as well
as for the proton fusion process.

\subsection*{Acknowledgments}

We are grateful for correspondence from Gerry Brown and John Bahcall
whose suggestions triggered part of this work.
One of us (TSP) is indebted to Professor 
Francisco Fern\'andez for the hospitality at
Grupo de F\' isica Nuclear, Universidad de Salamanca, where most of this
work was done and to Professor C.W. Kim for his hospitality at Korea 
Institute for Advanced Study where this paper was finalized. The work of 
KK was partially supported by the NSF Granty No. PHYS-9602000 and that
of DPM by the KOSEF through CTP of SNU and by the Korea Ministry of Education.

\newpage
\section*{Appendix}
\indent

We briefly comment on a recent discussion \cite{KSWPDS} on
Weinberg's counting rule. For related discussions on this issue,
see \cite{gegelia2,gegelia,steele,birse,beane,phillips}. More recently,
Birse, McGovern and Richardson~\cite{RGE} pointed out that for two-nucleon 
scattering with a quasi-bound
state nearby as involved in these discussions, fluctuations should be made 
around a non-trivial fixed point. We shall not address this 
issue here.

The authors of
\cite{KSWPDS} have argued that the counting rule eq.(\ref{nuChPT})
breaks down
when applied to {\it reducible} graphs,
which need to be summed to all orders
to describe bound or quasi-bound states.
If one were to use the usual dimensional regularization,
this infinite series with a finite set of irreducible graphs 
for a potential would require an infinite number of counter terms 
to regularize the series~\cite{lm}.

In constructing counting rules of low-energy effective theories
that are based on derivative expansion,
subtlety (if any) resides in loop integrals.
To be definite, we consider an $n$-dimensional integral
of a function $f(\bfp, \,\bfq)$,
where $\bfp$ stands for an external momentum
and $\bfq$ a loop-momentum,
\be
I^\Lambda(\bfp) \equiv \int^\Lambda d^n\bfq\, f(\bfp,\,\bfq),
\ee
where the superscript $\Lambda$
indicates that the quantity is regulated with a cutoff $\Lambda$.
Let $f(\bfp,\,\bfq)$ be of order of $Q^{\nu_f}$
when both $\bfp$ and $\bfq$ are of order of $Q$.
Weinberg assumed that the integral should be counted as
\be
I^\Lambda(\bfp) = \calO(Q^{n + \nu_f}).
\label{wein:assume}\ee
Note that we do not write down here explicitly the power of the
``large scale" $\Lambda$.
If one naively assumes
that $I^\Lambda(\bfp) \sim \Lambda^n Q^{\nu_f} \sim Q^{\nu_f}$,
where  $\sim$ stands for ``the same order",
then one would find that Feynman graphs with arbitrary numbers of loops
are not suppressed, which would make the effective theory
completely useless.
To elucidate Weinberg's assumption (\ref{wein:assume}),
let us expand $I^\Lambda(\bfp)$ in decreasing powers of $\Lambda$,
\be
I^\Lambda(\bfp) = \sum_{k=0}^{\infty} \Lambda^{n-k}
I_k^\Lambda(\bfp)
\label{Iexp}\ee
with
\be
I_k^\Lambda(\bfp) \sim Q^{\nu_f + k}.
\ee
The first $n$-terms ($0\le k < n$) -- which diverges when the cutoff
$\Lambda$ goes to infinity -- appear as (finite) polynomial in
$\bfp$. If the loop integral is for an irreducible loop,
then the polynomials are to be absorbed into the counter
terms of the corresponding vertex functions
in the renormalization procedure.
Then these terms become ``invisible" and hence do not
affect any physical observables.
This implies that, in constructing counting rules,
one can effectively count the cutoff of the loop as order of $Q$,
even though the ``actual cutoff" $\Lambda$ is much bigger than $Q$
and is not directly related to $Q$.
In our understanding, this feature leads to Weinberg's prescription:
Assign the ``effective cutoff", $\Lambda_*$,
\be
\Lambda_* \sim Q \ll \Lambda ,
\ee
to any loop integrals in {\it irreducible} diagrams.
It is then to be expected that Weinberg's assumption (\ref{wein:assume})
cannot be simply extended to reducible loops.
In fact, the assumption should not work for reducible loops
if a system is to have a (quasi) bound state (see below).
Another problem is that, for reducible loops,
it is not obvious how to absorb the polynomials,
$I_k(\bfp)$ with $k < n$, into counter terms.
The standard quantum field theory prescription tells us
only how to renormalize vertex functions which receive contributions
only from irreducible loops and counter terms.

To further expound these difficulties,
we examine the counting rules for {\it reducible} graphs.
The Lippman-Schwinger equation is schematically of the form,
$T = V + VG^0T$, where $V$ is a {\it regulated} potential,
$G^0$ the free two-nucleon propagator
and $T$ the $T$-matrix.
For a bound state, there is a pole in the $T$ matrix satisfying the
homogeneous equation
$T= VG^0T$, which implies $VG^0 = 1$.
For quasi-bound states, we have  $VG^0 = \calO(1)$.
For simplicity, let us retain only the $C_0$ (leading-order contact)
term and assume the separability of the potential.
Then the required condition for a reducible loop graph is
\be
C_0 \int^\Lambda d^3 \bfq\, \frac{q^2}{ME - q^2 + i0^+}
\sim
- C_0 \int^\Lambda d q
\sim 1\,.
\label{red}\ee
In finite cut-off effective theory, as noted in \cite{pkmr-prc},
$C_0$ scales as $\Lambda^{-1}$ and the integral
$\int^\Lambda d q \sim \Lambda$,
so the condition (\ref{red}) is trivially satisfied.
In the Weinberg counting rule (\ref{nuChPT}),
as we have surveyed,
every loop in $n$-dimensional space is counted as
$\calO(Q^n)$. When reducible graphs are present, we need to introduce
a modification to this rule.
We note that the following rule is consistent with eq.(\ref{red}):
Any {\it reducible} loop in $n$-dimensional space should be counted as
$\calO(Q^{n-1})$.
It is also interesting to find that eq.(\ref{red})
implies that the very natural scale of $|C_0|\sim \Lambda^{-1}$
forces {\it any} low-energy two-nucleon interaction to become non-perturbative
provided that $C_0<0$, which
is consistent with the shallow-binding of the deuteron
and the almost-binding of $np$ ${}^1S_0$ channel in nature.

In the dimensional regularization (DR) scheme, however,
realizing the condition (\ref{red}) is non-trivial
due to the {\it lemma} $\int d^n \bfq = 0$.
This could be a reason why  EFTs based on the naive DR fails
in describing low-energy ${}^1S_0$ scattering \cite{KSWPDS,lm}.
In \cite{KSWPDS}, a new counting rule was suggested
involving subtractions not only at $D=4$ but for
all $D \le 4$, where $D$ is the space-time dimension. In this scheme,
the condition corresponding to (\ref{red}) takes the form
\be
C_0 \mu \sim 1
\label{Dred}\ee
with
\be
C_0 = \frac{1}{1/a - \mu},
\label{C0PDS}\ee
where $\mu$ is the renormalization scale.
Eq.(\ref{C0PDS}) satisfies the corresponding
consistency condition eq.(\ref{Dred}),
which is responsible for the success of the PDS scheme
in contrast to the failure of the
naive dimmensional regularization.\footnote{\protect
One of the powers of the PDS scheme is the
ability to write an analytic expression for $C_0$
in terms of the {\em unnaturally} large scattering
length $a$, eq.(\ref{C0PDS}), reflecting a large 
anomalous dimensiion of the four-Fermi (baryon) interaction. 
This feature is somewhat buried in the cut-off method although
it is implicit in it.}
However, if we were to count the scale $\mu$ as of order of $Q$
as was done in \cite{KSWPDS},
there would appear an inconsistency
with the Georgi-Manohar counting rule \cite{georgi}
that $\frac{4\pi}{M} C_0 \sim \frac{1}{f_\pi^2} \sim \frac{1}{\Lambda^2}$.
On the other hand, if one were to adopt the PDS scheme but count
the renormalization scale as $\mu \sim \Lambda$ appropriate
for {\it local} operators,
then one would be led to $C_0 \sim \frac{1}{\Lambda}$
and hence to Weinberg's counting rule.

\pagebreak


\begin{thebibliography}{99}
\bibitem{gegelia2} J. Gegelia, ``Are pions perturbative in effective
field theory?", nucl-th/9806028.
\end{thebibliography}
\end{document}